# 3 kW passive-gain-enabled metalized Raman fiber amplifier based on passive gain

**Yizhu Chen, Tianfu Yao, Hu Xiao, Jinyong Leng, and Pu Zhou** [*]

College of Advanced Interdisciplinary Studies, National University of Defense Technology, Changsha 410073, China

[*]zhoupu203@163.com

**Abstract:** Raman fiber lasers (RFLs) are currently promising and versatile light sources for a variety of applications. So far, operations of high power and brightness-enhanced RFLs have absorbed enormous interests along with rapid progress. Nevertheless, the stable Raman lasing at high power levels remains challenged by the thermal effects. In an effort to realize more effective thermal management in high power RFLs, here we demonstrate, for the first time, an all-fiberized RFA employing metal-coated passive fiber enabling high power lasing. By employing aluminum to the cladding of graded-index (GRIN) passive fiber, the thermal abstraction of the laser devices is more sufficient to support low-temperature operation. The maximum output power reaches 3.083 kW at 1130 nm with a conversion efficiency of 78.7%. To the best of our knowledge, this is the first Raman laser generation based on metal-coated passive fiber. Meanwhile, it is also the highest power attained in the fields of all kinds of Raman lasers based on merely nonlinear gain.

1. Introduction

Stimulated Raman scattering (SRS), one of the inelastic interactions between light and particles, can generate frequency-converted lasing relative to the given pumping source [1-6]. In contrast to the well-known rare-earth (RE)-doped fiber lasers, RFLs employing SRS can not only reach those unattainable wavelengths with high power level, but also benefit from the easier thermal management [7], and absence of photodarkening [8]. For this appealing feature, the RFLs have been broadly exploited and implemented to support scalable power within the transmission window of silica fibers [7,9-18].

Furthermore, during the large interaction fiber length, RFLs can also generate brightness-enhanced laser corresponding to the pump laser with relatively low brightness [19]. Fig. 1 shows the power revolution of high power RFLs with BE mainly accomplished by two approaches. One approach is based on cladding pumping, where the pump light is launched into the inner clad and Stokes light emits from the core of the passive fiber [20-24]. However, the emergence of higher order Stokes hampers the power boosting of signal laser. To overcome this challenge, the cladding/core area ratio of Raman fiber is normally demanded to be less than 8 [25], which is much smaller than that of the commercial double-clad fiber. As the BE is related to the cladding/core area ratio, the pump brightness is required to be relatively high to obtain the Raman output with high beam quality. This imposes the cost of fabricating and the complexity of both pump and fiber designing.

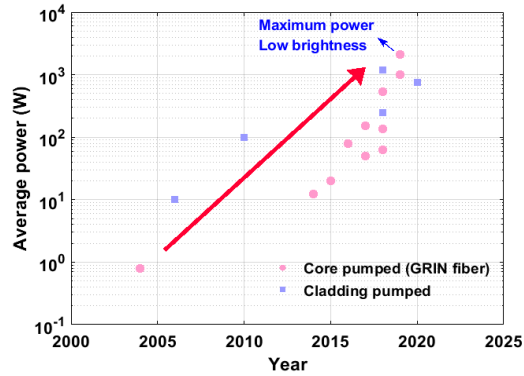

Fig. 1. Evolution of the average output power of high-power brightness-enhanced Raman laser (all-fiber or free-space configuration, resonator or amplifier, core pumped or cladding pumped, continuous wave or pulsed) over the past 20 years.

Rather than the cladding-pumping scheme, the core-pumped RFLs based on the standard GRIN passive fiber offers an economical way to perform BE [26,27]. In the core of GRIN fiber, the radial distribution of refractive-index is parabolic, where beam cleanup effect can provide improved output brightness. So far, much efforts have been devoted to RFLs employing GRIN fibers to support power scaling and BE simultaneously, as shown in Figure 1. The early Raman lasers with GRIN fiber were pumped by solid-state lasers, generating low-power but brightness-enhanced laser to validate the feasibility of BE in GRIN fibers [28,29]. The realization of Laser Diode (LD)-pumping brought about the rapid power ramping in GRIN Raman lasers. Nevertheless, the complex coupling and reflection of laser in the free-space configuration resulted in limitation of either the output efficiency or the beam quality [30-33]. The all-fiber configuration is accomplished by the manufacturing of fiberized combiner and fiber Bragg gratings (FBGs) based on GRIN fiber [34]. It motivated the reliability and robustness of GRIN RFLs with the power amounting to over 100 W [35-37]. However, to further scale the output power, the current available pump power from low-brightness LDs, and the high-power operation of the FBGs based on the thick GRIN fiber may both be the limitations.

As a substitutive approach, Raman fiber amplifier (RFA) pumped by combined fiber lasers and seeded by low power resonator, has shown great potential thanks to the significantly releasing on pump brightness and the special FBGs [24,38-42]. In our previous studies, we studied the all-fiber RFA pumped by fiber lasers and amplified the seed laser through a piece of GRIN fiber [40-42]. Based on progressively optimization, over 2 kW output power was achieved in the RFA with significant power enhancement compared with resonators [42]. The output beam quality improved at low power levels, whilst the beam quality and BE factor worsened at high power level with great fluctuation of output beam spot. In the 2-kW amplifier, we found it challenging to remove the heat load generated in the coiled Raman fiber and pigtailed passive fiber of end cap, resulting in serious thermal effects in the RFA at high power level. The similar thermal-induced instability associated with SRS has also been studied by other research groups [43-46], which not only influence the stable output performance of laser, but also would ultimately limit the power scaling and BE in RFLs. Thus, the approaches to the more efficient thermal management in the high-power Raman laser is urged.

In term of the thermal effects suppression, researchers on RE-doped fiber lasers have demonstrated several effective techniques, such as tandem pumping [47,48], metal-coating fiber [49,50] and so on. On the one hand, referring to tandem pumping scheme, the Raman generation is in agreement with low quantum defect to ease the heat management. On the other hand, the metal-coated active fibers have been applied in RE-doped fiber lasers, allowing the laser with high-temperature handling and improved threshold of thermal effects. The application of metal-coating enlightens us about the further optimization of RFA. Hence, in the current work, we demonstrate the all-fiberized RFA employing metal-coated passive fiber for the first time. By applying metal coating to a piece of GRIN passive fiber, the heat abstraction of the laser system is more sufficient, enabling the fiber with low-temperature operation. The maximum output power is 3.083 kW at 1130 nm with conversion efficiency of 78.7%. To the best of our knowledge, this is the highest power Raman lasers in any kind of configuration employing merely Raman gain, even higher than that obtained via RE-doped fiber lasers at this wavelength.

## 2. Experimental setup

The schematic of the high power RFA employing metal-coated GRIN fiber is depicted in Fig. 2(a). The amplifier applies master oscillator power amplifier (MOPA) configuration, which is composed of one seed oscillator stage and an amplification stage. A standard Raman fiber oscillator at 1130 nm serves as the seed laser, pumping by an Ytterbium-doped fiber laser (YDFL) at 1080 nm [42]. The maximum output power of the seed laser is ~200 W, and we experimentally confirmed a moderate seed laser power of 85 W to support efficient power amplification, including the signal laser at 1130 nm of 57.3 W as well as the unconverted pump laser of 27.7 W. In the amplification stage, the pump lasers are high power YDFLs at 1080 nm, enabling a maximum output power of ~4 kW. The tail fibers of pump and seed laser both have core/cladding diameters of 20 μm/400 μm, respectively.

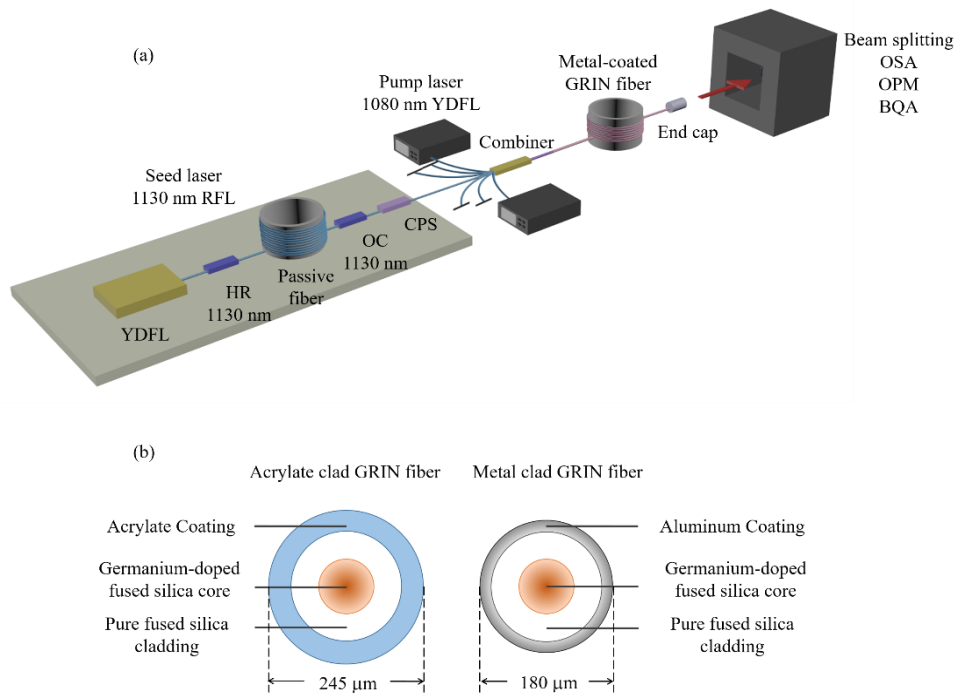

Fig. 2. (a) Experimental layout of the high power RFA based on metal-coated GRIN passive fiber. (b) The cross section of the common acrylate-coated GRIN fiber and the metal-coated GRIN fiber. HR:

High reflectivity FBG. OC: Output coupler FBG. CPS: Cladding pump stripper.

In order to meet the requirement of efficient heat elimination, a piece of metal-coated GRIN fiber is employed as the gain and stokes shifting medium, as is schematically shown in Fig. 2(b). The core and cladding of the fiber is Germanium-doped fused silica and pure fused silica, which are the same with the common GRIN multimode fiber. The core/cladding diameters of the GRIN fiber are 62.5 μm/125 μm, and the NA of fiber core is ~0.275. In the coating layer of the metal-coated fiber, the acrylate applied in the coating of the common GRIN fiber is replaced by aluminum with high thermal conduction, accompanied by the decreased diameter from ~245 μm to ~180 μm. The participation of metal in the coating and the thinning coating significantly increase the maximum operating temperature of the metal-coated fiber to over 400°C, which is much higher than the 85°C supported by the acrylate coating. Meanwhile, the metal coating is also helpful to better endurance of bending stress.

A home-made (6+1)×1 pump and signal combiner is utilized to couple the pump and seed lasers into the metal-coated passive fiber. All the input fibers of the combiner has identical core/cladding diameters to those of the tail fibers in pump and seed lasers. To preserve the power and beam quality after the launching of lasers, the tapered fused bundle of the combiner is optimally designed and manufactured. Then it is spliced with a piece of common GRIN fiber, whose radial parameters are the same with the metal-coated fiber, but with ordinary acrylate coating. Before the experiment, the performance of the pump and signal combiner is tested. The transmission efficiency of the combiner surpasses 95%, and the seeding beam quality after power combination is also well preserved.

After the optimization for power scaling, the total length of the GRIN fiber in the amplifier is ~20 m, including the output fiber of the combiner and the gain fiber. Additionally, a home-manufactured end cap is fusion spliced at the end of the metal-coated GRIN fiber to suppress backward lasing. For heat abstraction, the metal-coated GRIN fiber is coiled onto a metal column with diameter of 0.24 m. Water cooling plates maintaining at 20 °C are also employed to other fiber devices.

3. **Experimental result and discussions**

To measure the laser power as well as the beam quality of the amplifier, a beam splitting and detecting system is applied after the end cap. It is similar to the implementation in Ref. [42], whilst the beam collimating and splitting section applies more lens to attenuate and sample the laser due to the improvement of power level. By employing Optical power meter (OPM) with high-power endurance, the output power of the RFA is recorded and graphically presented in Fig. 3 as a function of launched pump power. With the incremental launching pump power, the total output power increases almost linearly. The signal laser power grows rapidly with enhanced Stokes conversion. For the maximum pump power of 3920 W, we obtain the total output power of 3535 W, with the corresponding signal laser power of 3083 W. At maximum power, the power ratio of signal laser to total output laser is 87.2%, and the conversion efficiency corresponding to the launched pump power is 78.7%. The power of residual pump laser increases firstly due to the inadequate conversion, then decreases gradually to 421.5 W at maximum output power. The 2$^{nd}$ order Raman laser centered at ~1184 nm arises when the pump power is 2915 W, and the power rises slowly to 30.3 W at maximum.

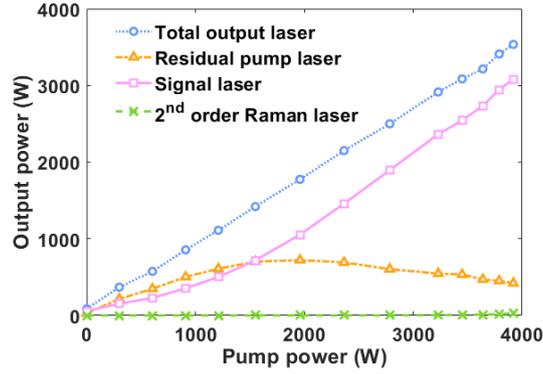

Fig. 3. The output power with various launched pump power, including total output power, residual pump power at 1080 nm, signal power at 1130 nm and higher order Raman laser power.

The output spectrum of the RFA is measured by Optical spectrum analyzer (OSA) and described in Fig. 4 with various signal power. In the seed laser, there is only signal laser at 1130 nm and a portion of residual pump laser at 1080 nm. With further incremental pump laser, the peak intensity of signal laser increases constantly. The 3-dB bandwidth of the seed laser at 1130 nm is ~1.52 nm, which increases with ramping power and ultimately reaches ~4.08 nm. In contrast to the maximal bandwidth of 2.12 nm in the previous 2-kW RFA [42], the spectrum broadens ulteriorly. It comes from the Four-wave-mixing and Cross-phase modulation in the long passive fiber with subtle peaks arising beside the signal spectrum [51,52]. Meanwhile, the individual pump laser spectrum broadens with higher power level, thus contributes to the wide signal laser spectrum through Raman conversion. The 2$^{nd}$ order Raman light is captured when the signal power is 353 W and increases gradually. At maximum output power, the peak intensity of 2$^{nd}$ order Raman laser is ~22 dB lower than the signal laser. The spectrum centered at 1035 nm appears almost simultaneously with the 2$^{nd}$ order Raman laser with a slow increasing tendency. The similar spectrum component has been observed in our previous research, and it results from the Four-wave-mixing in the long passive fiber [2].

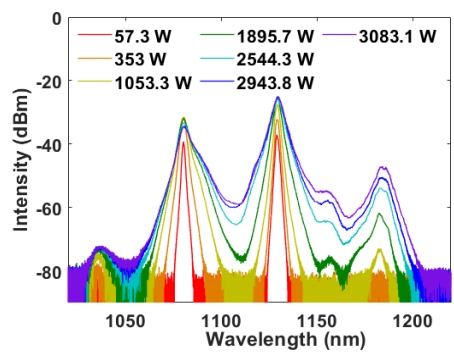

Fig. 4. The output spectrum of the RFA with varied signal laser power.

Through the optimal design of the metalized amplifier, we accomplish the highest power in the fields of all-passive-gain Raman lasers. To achieve power scaling of amplifier, high power pump laser is added to this amplifier. According to the definite upper limit of pump power, the seed laser power and the fiber length are both adjusted to better balance the requirement of higher 2$^{nd}$ order Raman threshold and efficient Raman conversion of signal laser.

Moreover, the application of metal-coating enables the Stokes shifting in the low-temperature environment, thus enhances the robustness of fiber amplifier and alleviates the thermal load at high power levels. Figure 5 shows the thermal image of the coiled metalized fiber on the metal column at maximum power. The whole piece of metal-coated fiber remains low temperature (lower than 28.2°C), whilst the highest temperature lies in the nonmetallic fixed point of the fiber (37.5°C in Fig. 5). As a consequence, the beam shape of laser is more stable than the previous result [42], and better beam quality as well as more effective BE are achieved in this 3-kW amplifier. Nevertheless, the temperature of the fiber cannot be completely controlled. As mentioned above, the thermal effect of the lenses influence the stable measurement of the beam quality. In addition, the pigtail fiber of the combiner is still ordinary acrylate-coated GRIN fiber, and the fusion points between the combiner, the metal-coated GRIN fiber and the end cap get rid of the metal-coating for splicing. The temperatures of these positions are quite high at maximum output power, resulting in the worsening of output beam quality as well. In future higher-power experiment, the all-metallization of fiber laser should be in consideration for more sufficient heat management.

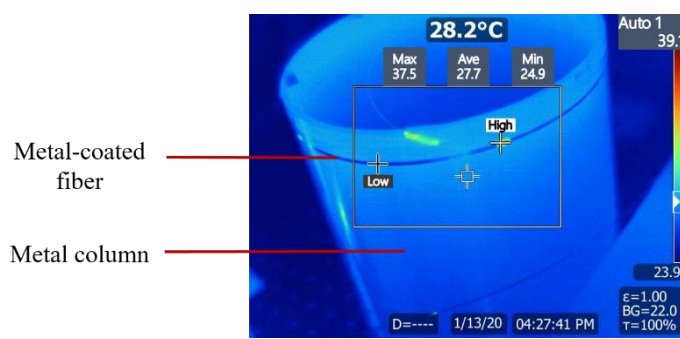

Fig. 5. The thermal image of the coiled metal-coated fiber on the column at maximum output power.

Further, the customized pump and signal combiner pigtailed with GRIN fiber is an important device in this amplifier. It well preserves the seeding beam quality after power combination, which is far better than the previous result of ~15.8 in [42]. The application of pigtailed GRIN fiber also eliminates the mode mismatching between the combiner and the gain fiber. It optimizes the combination and transmission of launched seed laser, and enables the amplification of high-brightness signal laser in the amplifier with higher BE. One can conclude that the beam quality of generated laser is related to the characteristics of seed laser and can be modulate for better output performance. With these efforts, we achieve the highest power among Raman lasers in any kind of configuration that merely based on Stokes gain. The power level at this wavelength is hardly obtained through RE-doped fiber lasers owing to the limited emission spectrum, which shows the superiority of the agile Raman lasers. In future experiment for more precise measurement, the lenses in the beam splitting system can be optimized and equipped with cooling system entirely for heat abstraction. Moreover, to realize further metallization of the amplifier, the metal-coated fiber can be utilized in the manufacturing of combiner and end cap, which will be beneficial to the power scaling of RFA.

## 4. Conclusions

To conclude, we demonstrate a high power all-fiberized Raman amplifier employing metal-coated fiber for the first time. To achieve this, a piece of metal-coated GRIN passive fiber is applied as the gain medium, enabling better thermal abstraction and low-temperature operation during the whole amplification process. The maximum output power at 1130 nm is up to 3.083 kW with conversion efficiency of 78.7%. To the best of our knowledge, this is the highest power among Raman lasers in any kind of configuration that merely based on Stokes gain. At this operating wavelength, the power level is even higher than that obtained through RE-doped fiber lasers.

**Reference**

1. R. H. Stolen, E. P. Ippen, and A. R. Tynes, "Raman Oscillation in Glass Optical Waveguide," Appl. Phys. Lett., vol. 20, no. 2, pp. 62-64, 1972.
2. G. P. Agrawal, Nonlinear Fiber Optics (Fourth Edition) (Academic Press, 2007).
3. V. R. Supradeepa, Y. Feng, and J. Nicholson, "Raman fiber lasers", J. Optics-Uk, 19, 23001(2017).
4. Y. Feng, Raman Fiber Lasers (Springer, 2017).
5. R. Casula, J. Penttinen, M. Guina, A. J. Kemp, and J. E. Hastie, "Cascaded crystalline Raman lasers for extended wavelength coverage: continuous-wave, third-Stokes operation," Optica 5(11), 1406-1413 (2018).
6. X. Yang, O. Kitzler, D. Spence, Z. Bai, Y. Feng, and R. Mildren, "Diamond Sodium Guide Star Laser," Opt Lett 45(7), 1898-1901 (2020).
7. T. Yao, and J. Nilsson, "835 nm fiber Raman laser pulse pumped by a multimode laser diode at 806 nm," Journal of the Optical Society of America B 31(4), 882-888 (2014).
8. K. E. Mattsson, "Photo darkening of rare earth doped silica," Opt Express 19(21), 19797-19812 (2011).
9. Z. Wang, H. Wu, M. Fan, Y. Rao, X. Jia, and W. Zhang, "Third-order random lasing via Raman gain and Rayleigh feedback within a half-open cavity," Opt Express 21(17), 20090-20095 (2013).
10. J. Liu, D. Shen, H. Huang, C. Zhao, X. Zhang, and D. Fan, "High-power and highly efficient operation of wavelength-tunable Raman fiber lasers based on volume Bragg gratings," Opt. Express, vol. 22, no. 6, pp. 6605-6612, 2014.
11. V. R. Supradeepa, and J. W. Nicholson, "Power scaling of high-efficiency 1.5 μm cascaded Raman fiber lasers," Opt Lett 38(14), 2538-2541 (2013).
12. S. I. Kablukov, E. I. Dontsova, E. A. Zlobina, I. N. Nemov, A. A. Vlasov, and S. A. Babin, "An LD-pumped Raman fiber laser operating below 1 μm," Laser Phys. Lett., vol. 10, no. 8, pp. 85103, 2013.
13. H. Jiang, L. Zhang, and Y. Feng, "Silica-based fiber Raman laser at > 2.4 μm," Opt Lett 40(14), 3249-3252 (2015).
14. J. Liu, F. Tan, H. Shi, and P. Wang, "High-power Silica-based Raman Fiber Amplifier at 2147 nm," in *CLEO: 2015*(Optical Society of America, San Jose, California, 2015), pp. u5A-u39A.
15. T. Yin, Z. Qi, F. Chen, Y. Song, and S. He, "High peak-power and narrow-linewidth all-fiber Raman nanosecond laser in 1.65 μm waveband," Opt. Express, vol. 28, no. 5, pp. 7175-7181, 2020.
16. J. Song, H. Wu, J. Ye, H. Zhang, J. Xu, P. Zhou, and Z. Liu, "Investigation on extreme frequency shift in silica fiber-based high-power Raman fiber laser," High Power Laser Sci 6(2), 21-28 (2018).

17. S. Loranger, A. Tehranchi, H. Winful, and R. Kashyap, "Realization and optimization of phase-shifted distributed feedback fiber Bragg grating Raman lasers," Optica 5(3), 295-302 (2018).
18. Z. Wang, Q. Xiao, Y. Huang, J. Tian, D. Li, P. Yan, and M. Gong, "Dual-wavelength bidirectional pumped high-power Raman fiber laser," High Power Laser Sci 7(e5) (2019).
19. Y. Glick, Y. Shamir, Y. Sintov, S. Goldring, and S. Pearl, "Brightness enhancement with Raman fiber lasers and amplifiers using multi-mode or multi-clad fibers," Opt. Fiber Technol., vol. 52, no., pp. 101955, 2019.
20. C. A. Codemard, J. K. Sahu, and J. Nilsson, "100-W CW cladding-pumped Raman fiber laser at 1120 nm," Proceedings of SPIE - The International Society for Optical Engineering 7580(6), 491-513 (2010).
21. T. Yao, A. V. Harish, J. K. Sahu, and J. Nilsson, "High-Power Continuous-Wave Directly-Diode-Pumped Fiber Raman Lasers," Applied Sciences 5(4), 1323-1336 (2015).
22. Y. Shamir, Y. Glick, M. Aviel, A. Attias, and S. Pearl, "250 W clad pumped Raman all-fiber laser with brightness enhancement.," Opt Lett 43(4), 711 (2018).
23. Y. Glick, Y. Shamir, M. Aviel, Y. Sintov, S. Goldring, N. Shafir, and S. Pearl, "1.2 kW clad pumped Raman all-passive-fiber laser with brightness enhancement," Opt Lett 43(19), 4755-4758 (2018).
24. Y. Chen, T. Yao, H. Xiao, J. Leng, and P. Zhou, "High-power cladding pumped Raman fiber amplifier with a record beam quality," Opt Lett 45(8), 2020.
25. J. Ji, C. A. Codemard, M. Ibsen, J. K. Sahu, and J. Nilsson, "Analysis of the Conversion to the First Stokes in Cladding-Pumped Fiber Raman Amplifiers," IEEE J Sel Top Quant 15(1), 129-139 (2009).
26. A. Polley, and S. E. Ralph, "Raman Amplification in Multimode Fiber," IEEE Photonic Tech L 19(4), 218-220 (2007).
27. N. B. Terry, T. G. Alley, and T. H. Russell, "An explanation of SRS beam cleanup in graded-index fibers and the absence of SRS beam cleanup in step-index fibers," *Opt. Express*, vol. 15, no. 26, pp. 17509-17519, 2007.
28. S. H. Baek, and W. B. Roh, "Single-mode Raman fiber laser based on a multimode fiber," Opt. Lett., vol. 29, no. 2, pp. 153-155, 2004.
29. B. M. Flusche, T. G. Alley, T. H. Russell, and W. B. Roh, "Multi-port beam combination and cleanup in large multimode fiber using stimulated Raman scattering," Opt. Express, vol. 14, no. 24, pp. 11748-11755, 2006.
30. T. Yao, A. V. Harish, J. K. Sahu, and J. Nilsson, "High-Power Continuous-Wave Directly-Diode-Pumped Fiber Raman Lasers," *Applied Sciences*, vol. 5, no. 4, pp. 1323-1336, 2015.
31. Y. Glick, V. Fromzel, J. Zhang, A. Dahan, N. Ter-Gabrielyan, R. K. Pattnaik, and M. Dubinskii, "High power, high efficiency diode pumped Raman fiber laser," Laser Phys Lett 13(6), 65101 (2016).
32. Y. Glick, V. Fromzel, J. Zhang, N. Ter-Gabrielyan, and M. Dubinskii, "High-efficiency, 154 W CW, diode-pumped Raman fiber laser with brightness enhancement," Appl Optics 56(3), B97 (2017).
33. E. A. Zlobina, S. I. Kablukov, A. A. Wolf, A. V. Dostovalov, and S. A. Babin, "Nearly single-mode Raman lasing at 954 nm in a graded-index fiber directly pumped by a multimode laser diode," *Opt. Lett.*, vol. 42, no. 1, pp. 9-12, 2017.
34. S. I. Kablukov, E. A. Zlobina, M. I. Skvortsov, I. N. Nemov, A. A. Wolf, A. V. Dostovalov, and S. A. Babin, "Mode selection in a directly diode-pumped Raman fibre laser using FBGs in a graded-index multimode fibre," Quantum Electron+ 46(12), 1106-1109 (2016).
35. E. A. Zlobina, S. I. Kablukov, A. A. Wolf, I. N. Nemov, A. V. Dostovalov, V. A. Tyrtyshnyy, D. V. Myasnikov, and S. A. Babin, "Generating high-quality beam in a multimode LD-pumped all-fiber Raman laser," Opt. Express, vol. 25, no. 11, pp. 12581-12587, 2017.
36. E. A. Evmenova, S. I. Kablukov, I. N. Nemov, A. A. Wolf, A. V. Dostovalov, V. A. Tyrtyshnyy, D. V. Myasnikov, and S. A. Babin, "High-efficiency LD-pumped all-fiber Raman laser based on a 100 μm core graded-index fiber," Laser Phys Lett 15(9), 95101 (2018).
37. Y. Glick, Y. Shamir, A. A. Wolf, A. V. Dostovalov, S. A. Babin, and S. Pearl, "Highly efficient all-fiber continuous-wave Raman graded-index fiber laser pumped by a fiber laser.," Opt Lett 43(5), 1027-1030 (2018).
38. A. K. Sridharan, J. E. Heebner, M. J. Messerly, J. W. Dawson, R. J. Beach, and C. P. J. Barty, "Brightness enhancement in a high-peak-power cladding-pumped Raman fiber amplifier," Opt. Lett., vol. 34, no. 14, pp. 2234-2236, 2009.
39. W. Wang, L. Huang, J. Leng, S. Guo, and Z. Jiang, "Beam cleanup of the stimulated Raman scattering in grade-index multi-mode fiber," Chin Opt Lett, vol. 12, no. s2, pp. S21401, 2014.
40. Y. Chen, J. Leng, H. Xiao, T. Yao, and P. Zhou, "High-efficiency all-fiber Raman fiber amplifier with record output power," Laser Phys Lett 15(8), 85104 (2018).
41. Y. Chen, J. Leng, H. Xiao, T. Yao, and P. Zhou, "Pure Passive Fiber Enabled Highly Efficient Raman Fiber Amplifier with Record Kilowatt Power," IEEE Access 7, 28334-28339 (2019).
42. Y. Chen, T. Yao, L. Huang, H. Xiao, J. Leng, and P. Zhou, "2 kW high-efficiency Raman fiber amplifier based on passive fiber with dynamic analysis on beam cleanup and fluctuation," Opt Express 28(3), 3495-3504 (2020).
43. L. G. Wright, Z. Liu, D. A. Nolan, M. J. Li, D. N. Christodoulides, and F. W. Wise, "Self-organized instability in graded-index multimode fibres," Nat Photonics (2016).
44. S. Naderi, I. Dajani, J. Grosek, and T. Madden, "Theoretical treatment of modal instability in high-power cladding-pumped Raman amplifiers," Proceedings of SPIE - The International Society for Optical Engineering 9344, 93442X (2015).
45. Q. Chu, Q. Shu, Z. Chen, F. Li, D. Yan, C. Guo, H. Lin, J. Wang, F. Jing, C. Tang, and R. Tao, "Experimental study of mode distortion induced by stimulated Raman scattering in high-power fiber amplifiers," Photonics Res 8(4), 595-600 (2020).
46. D. Victor, M. Friedrich, S. Maximilian, P. Gonzalo, W. Till, S. Thomas, and T. Andreas, "High power narrow-linewidth Raman amplifier and its limitation," 11260, 1126005 (2020).
47. H. Xiao, J. Leng, H. Zhang, L. Huang, J. Xu, and P. Zhou, "High-power 1018 nm ytterbium-doped fiber laser and its application in tandem pump," Appl. Opt. 54, 8166–8169 (2015).
48. P. Zhou, H. Xiao, J. Leng, J. Xu, Z. Chen, H. Zhang, and Z. Liu, "High-power fiber lasers based on tandem pumping," Journal of the Optical Society of America B 34(3), A29-A36 (2017).
49. C. X. Yu, O. Shatrovoy, T. Y. Fan, and T. F. Taunay, "Diode-pumped narrow linewidth

multi-kilowatt metalized Yb fiber amplifier," Opt Lett 41(22), 5202-5205 (2016).
50. J. M. O. Daniel, N. Simakov, A. Hemming, W. A. Clarkson, and J. Haub, "Metal clad active fibres for power scaling and thermal management at kW power levels," Opt Express 24(16), 18592-18606 (2016).
51. S. A. Babin, D. V. Churkin, A. E. Ismagulov, S. I. Kablukov, and E. V. Podivilov, "Spectral broadening in Raman fiber lasers," Opt Lett 31(20), 3007-3009 (2006).
52. G. Ravet, A. A. Fotiadi, and P. Mégret, "Spectrum broadening in Raman fiber laser induced by cross-phase modulation," in CLEO/Europe and IQEC 2007 Conference Digest (Optical Society of America, Munich, 2007), J4-J7.